\documentclass[14pt, a4paper]{article}

\usepackage{arxiv}
\usepackage{amsmath}
\usepackage{amssymb}
\usepackage{amsfonts}
\usepackage{graphicx}
\usepackage{tikz}
\usepackage{circuitikz}
\usepackage{pgfplots}
\usepackage{geometry}
\usepackage{booktabs}
\usepackage{setspace}
\usepackage{caption}
\usepackage{subcaption}

\pgfplotsset{compat=1.18}

\title{Electromagnetic Characterization of magnetic ring: Case of square cross section shape}

\author{%
  Taha El Hajji*,\quad Lars Sjöberg \\
  Alvier Mechatronics AB, Helsingborg, Sweden  \\
  \texttt{taha.elhajji@gmail.com}
}

\begin{document}

\maketitle

\begin{spacing}{1.8}
\begin{abstract}
This paper presents a comprehensive 2D analytical model of a toroidal magnetic ring with a square cross-section, subjected to sinusoidal excitation. By applying Maxwell's equations in local Cartesian coordinates and utilizing a complex permeability framework, the exact analytical expressions for the internal magnetic field, flux, complex impedance, and losses are derived. The model rigorously separates eddy current losses, hysteresis losses, and winding losses, explicitly accounting for the skin effect and complex permeability within the conductive core using separation of variables and hyperbolic functions. Furthermore, parameter for apparent permeability is expressed to map the core behavior onto simplified linear material models. The derivations establish a mathematical foundation highly suitable for standardized material characterizations, such as Brockhaus and Iwatsu ring measurements, by avoiding the heavy computational cost of 2D and 3D Finite Element Analysis.
\end{abstract}
\end{spacing}
\keywords{Toroidal Magnetic Ring \and Eddy Currents \and Hysteresis \and Complex Permeability \and Maxwell's Equations}

\clearpage

\section{Introduction}
Accurate characterization of soft magnetic materials is vital for optimizing high-efficiency electromagnetic devices. Traditional measurement techniques, such as those performed using Brockhaus testers or Iwatsu systems, rely on extracting material properties by winding the specimen as a toroidal ring and measuring its voltage-current responses. However, as industry demands push excitation frequencies from 50Hz into the MHz range, the standardization of these measurement frameworks faces significant challenges regarding accuracy.
At high frequencies, the classical assumption of a uniform magnetic flux distribution across the core cross-section becomes invalid. The skin effect, driven by eddy currents within the conductive magnetic core, forces the flux toward the surface. In a core with a square cross-section, this effect leads to significant flux crowding at the edges and corners, creating a highly non-uniform field profile that empirical models fail to capture \cite{ieee4527040}. Furthermore, the material exhibits hysteresis, which introduces a phase lag between the applied field and the resulting flux. Modern characterization now necessitates a complex permeability framework to accurately map these frequency-dependent energy storage and loss mechanisms \cite{sd_ferrite}.
Historically, engineers relied on empirical fits like the Steinmetz equation \cite{ieee9003248}, which offer no physical insight into internal spatial fields or the shielding effects caused by the core geometry. While Finite Element Analysis (FEA) can model these distributions, the computational cost for iterative material characterization is often prohibitive. Consequently, there is a renewed focus on developing exact analytical solutions for toroidal geometries to calculate eddy current distributions and volume-averaged losses \cite{ieee668058, ieee5382736}.
The motivation of this work is to provide a rigorous 2D analytical model of the complex impedance that accounts for the simultaneous impact of eddy currents and hysteresis in the case of ring with square cross section. By solving Maxwell's equations in a 2D local Cartesian coordinate system applied to the toroid's square cross-section, this paper provides exact analytical solutions for the fields, flux, impedance, and losses. Utilizing a complex permeability model allows for the explicit mathematical separation of hysteresis and eddy current losses, a distinction critical for advanced core loss predictions and standardized material testing.

\section{Nomenclature}
\begin{itemize}
    \item $D_m$: Mean diameter of the toroidal ring [m]
    \item $ID$: Inner diameter of the toroidal ring [m]
    \item $OD$: Outer diameter of the toroidal ring [m]
    \item $a$: Side length of the square cross-section, where $a = (OD - ID)/2$ [m]
    \item $A$: Cross-sectional area, calculated as $A = a^2$ [m$^2$]
    \item $l_m$: Mean magnetic path length, evaluated as $l_m = \pi D_m$ [m]
    \item $\mu_c$: Complex magnetic permeability [H/m]. Expressed in rectangular form as $\mu_c = \mu' - j\mu''$ and in polar form as $\mu_c = |\mu_c| e^{-j\theta}$
    \item $\mu'$: Real part of complex permeability representing energy storage [H/m]
    \item $\mu''$: Imaginary part of complex permeability representing energy loss [H/m]
    \item $|\mu_c|$: Magnitude of complex permeability, defined as $|\mu_c| = \sqrt{\mu'^2 + \mu''^2}$ [H/m]
    \item $\theta$: Hysteresis loss angle [rad]
    \item $\sigma$: Electrical conductivity of the core material [S/m]
    \item $\omega$: Angular frequency of the sinusoidal excitation, $\omega = 2\pi f$ [rad/s]
    \item $N_1$: Number of turns of the primary winding
    \item $N_2$: Number of turns of the secondary winding
    \item $I$: RMS current supplied to the primary winding [A]
    \item $V_1, V_2$: Voltage across primary and secondary windings [V]
    \item $E_1$: Induced electromotive force (EMF) [V]
    \item $\mathbf{H}$: Magnetic field intensity [A/m]
    \item $\mathbf{B}$: Magnetic flux density [T]
    \item $\mathbf{E}$: Electric field intensity [V/m]
    \item $\mathbf{J}$: Current density [A/m$^2$]
    \item $\mathbf{D}$: Electric flux density [C/m$^2$]
    \item $\rho_v$: Volume charge density [C/m$^3$]
    \item $R_{DC}$: DC resistance of the primary winding [$\Omega$]
    \item $R_{winding}$: AC resistance of the primary winding accounting for skin effect [$\Omega$]
    \item $Z_{core}$: Complex impedance of the core [$\Omega$]
    \item $Z$: Total complex impedance of the primary circuit [$\Omega$]
    \item $Y_{core}$: Complex admittance of the core, defined as $Y_{core} = G_{core} - jB_{core}$ [S]
    \item $G_e$: Eddy current equivalent conductance [S]
    \item $G_h$: Hysteresis equivalent conductance [S]
    \item $P_e$: Eddy current power loss [W]
    \item $P_h$: Hysteresis power loss [W]
    \item $P_w$: Winding power loss [W]
    \item $x, y, z$: Local Cartesian coordinates within the square cross-section [m]
    \item $\gamma$: Complex propagation constant, defined as $\gamma = \sqrt{j\omega \mu_c \sigma}$ [1/m]
    \item $k$: Complex separation constant in the core, evaluated as $k = \sqrt{j\omega \mu_c \sigma / 2}$ [1/m]
    \item $\delta_w$: Skin depth in the winding wire, $\delta_w = \sqrt{2/(\omega \mu_w \sigma_w)}$ [m]
    \item $\xi$: Dimensionless size factor for the winding wire, $\xi = \sqrt{2} r_w / \delta_w$
    \item $r_w$: Radius of the primary winding wire [m]
    \item $\sigma_w$: Electrical conductivity of the winding wire [S/m]
    \item $\mu_w$: Magnetic permeability of the winding wire [H/m]
    \item $\mu_{app}$: Apparent complex permeability accounting for eddy currents [H/m]
    \item $\sigma_{app}$: Apparent electrical conductivity characterizing total core losses [S/m]
    \item $k_{app}$: Wave number using apparent conductivity, $k_{app} = \sqrt{j\omega \mu' \sigma_{app} / 2}$ [1/m]
    \item $\cosh, \sinh$: Hyperbolic cosine and sine functions
    \item $J_0, J_1$: Bessel functions of the first kind of zero and first order
    \item $\text{ber}, \text{bei}$: Kelvin functions of the first kind
\end{itemize}

\section{Modeled Ring}

\subsection{3D Representation of the Toroidal Ring}
\begin{figure}[h!]
\centering
\includegraphics[width=0.5\textwidth]{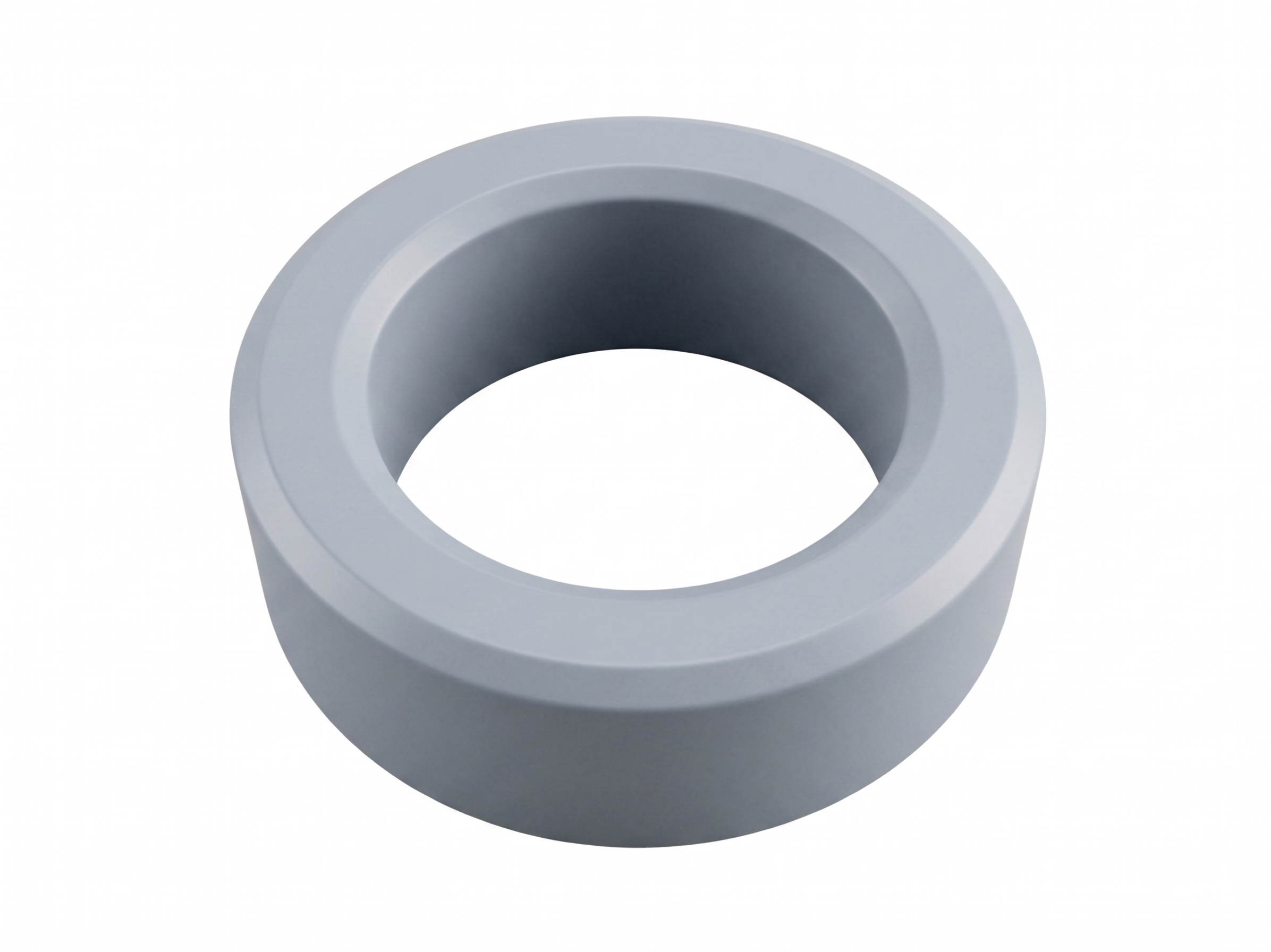}
\caption{3D visualization of the Toroidal Ring}
\label{ring3D}
\end{figure}

Fig. \ref{ring3D} represents a 3D visualization of the ring under study, a toroidal ring with a square cross-section.

\begin{figure}[htbp]
    \centering
    \begin{subfigure}{0.45\textwidth}
        \centering
        \includegraphics[width=\textwidth]{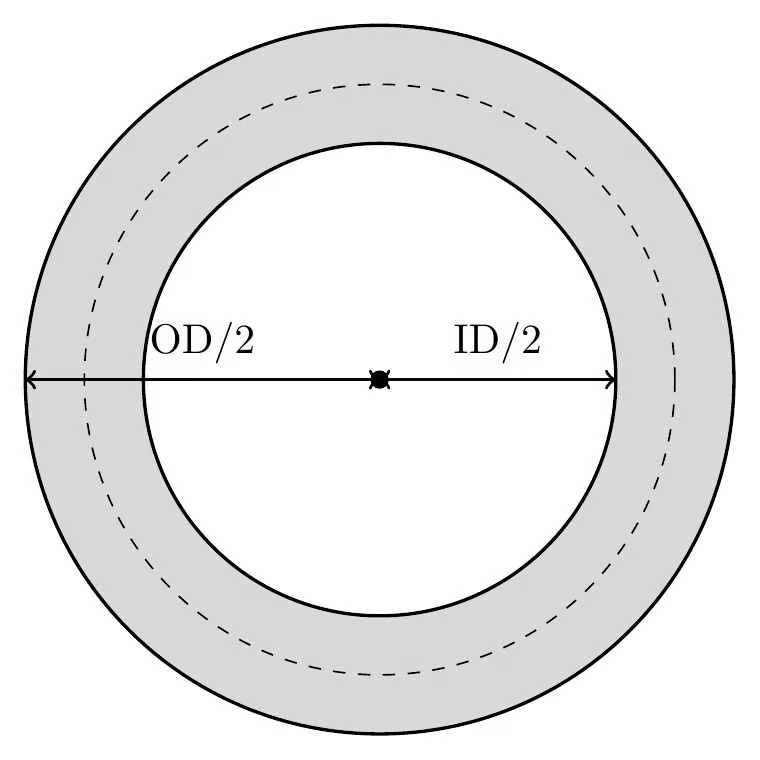}
        \caption{}
        \label{fig:left}
    \end{subfigure}
    \hfill
    \begin{subfigure}{0.4\textwidth}
        \centering
        \includegraphics[width=\textwidth]{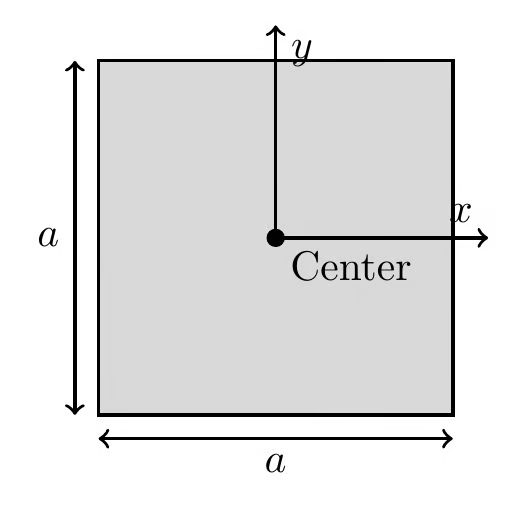}
        \caption{}
        \label{fig:right}
    \end{subfigure}
    \caption{Visualization of the Toroidal Ring: (a) Top cross section, (b) Side Cross-section}
    \label{fig:both}
\end{figure}

\subsection{Winding Configuration (Brockhaus Standard)}
\begin{figure}[h!]
\centering
\includegraphics[width=0.5\textwidth]{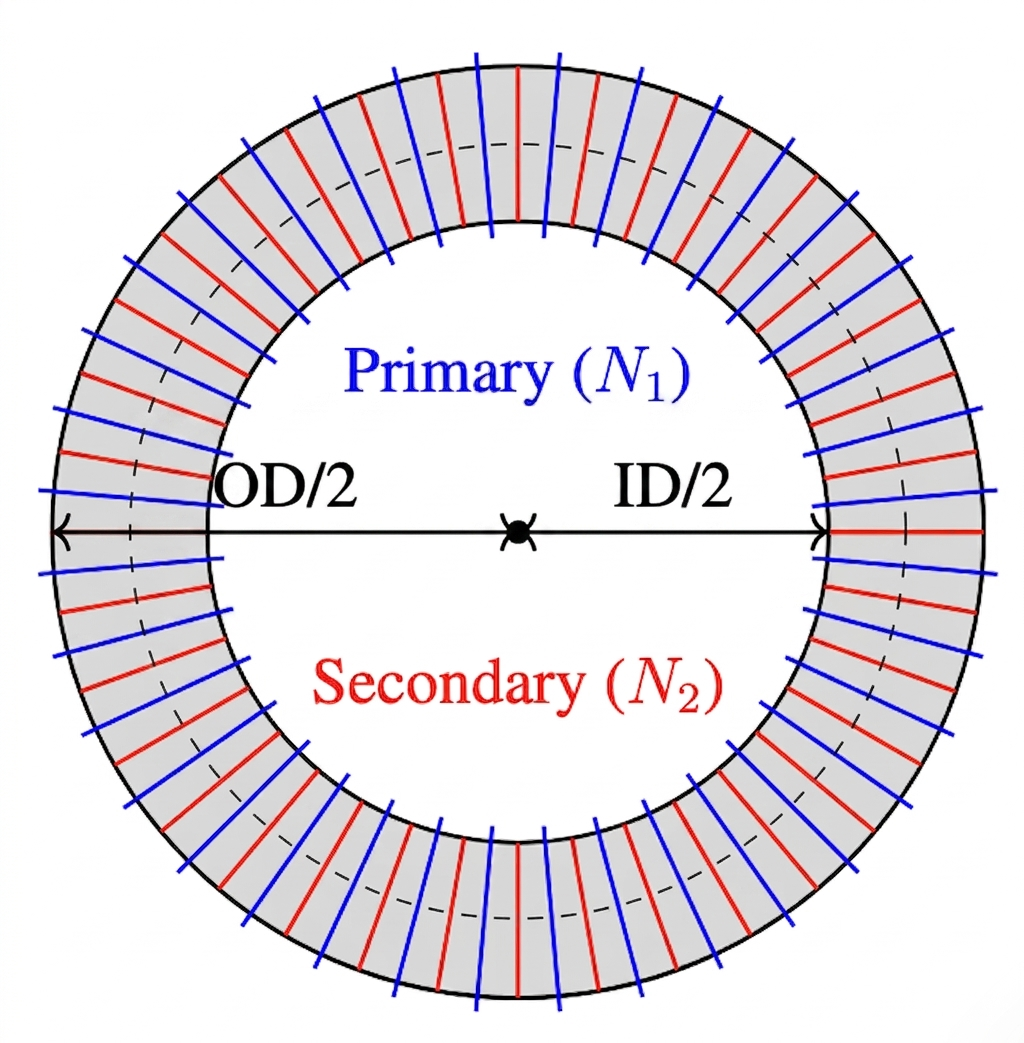}
\caption{Winding Arrangement}
\label{ringwinding}
\end{figure}

The dimensions of the ring are represented in fig.\ref{fig:both}. In standard material testing, the secondary winding ($N_2$) is wound directly onto the core to capture the true flux linkage. Leakage is minimized to negligible levels. The primary winding ($N_1$) is wound uniformly over the secondary. Fig. \ref{ringwinding} shows the winding configuration in 2D view.

\subsection{Equivalent Circuit Model}
\begin{figure}
    \centering
    \includegraphics[width=0.99\linewidth]{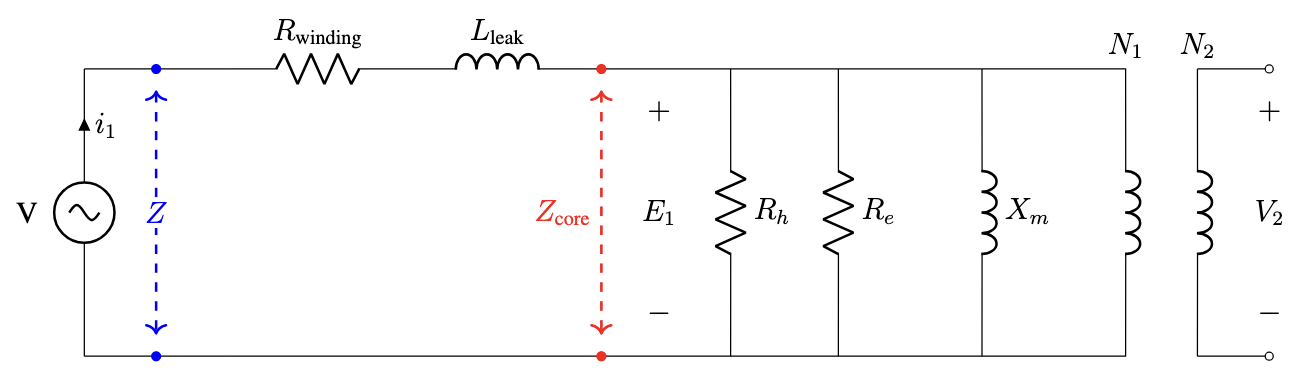}
    \caption{Equivalent magnetic circuit of the wouned ring}
    \label{fig:magcircuit}
\end{figure}

The electrical behavior of the system is represented in fig. \ref{fig:magcircuit}. The primary voltage $V_1$ consists of the voltage drop across the winding resistance, a residual leakage inductance $L_{leak}$, and the induced voltage $E_1$ at the core level. The core's non-ideal behaviors are represented by parallel branches for hysteresis ($R_h$), eddy currents ($R_e$), and magnetizing reactance ($X_m$). 

The governing equations for the transformer circuit are given as:
\begin{align}
    V_1 &= R_{winding} i_1 + L_{leak} \frac{di_1}{dt} + E_1 \\
    \frac{E_1}{N_1} &= \frac{V_2}{N_2}
\end{align}

\section{Modeling and Mathematical Proof}

\subsection{Fundamental Maxwell's Equations}
The derivation of the internal fields begins with the generalized Maxwell's equations:
\begin{align}
    \nabla \cdot \mathbf{D} &= \rho_v \quad \text{(Gauss's Law for Electricity)} \\
    \nabla \cdot \mathbf{B} &= 0 \quad \text{(Gauss's Law for Magnetism)} \\
    \nabla \times \mathbf{E} &= -\frac{\partial \mathbf{B}}{\partial t} \quad \text{(Faraday's Law of Induction)} \\
    \nabla \times \mathbf{H} &= \mathbf{J} + \frac{\partial \mathbf{D}}{\partial t} \quad \text{(Ampère's Circuital Law)}
\end{align}
In this model, since the quasi-static approximation is highly valid for the operating frequencies under consideration, the displacement current density is neglected. Thus, the analysis will primarily rely on Faraday's Law and the simplified Ampère's Law ($\nabla \times \mathbf{H} = \mathbf{J}$). Furthermore, all mathematical modeling and calculations in this study are performed at a no-load condition (the secondary circuit is an open circuit).

\subsection{Assumptions and Operational Constraints}
The analytical framework operates under the following key assumptions:
\begin{itemize}
    \item $OD/ID \le 1.2$: This assumption guarantees a uniform magnetic field in the DC case and at low frequency.
    \item $D_m>10.a$: this assumption is to ensure the equivalence between the 2D and 3D model.
    \item The core material features a uniform complex permeability $\mu_c = |\mu_c| e^{-j\theta}$ and a constant electrical conductivity $\sigma$.
    \item The core remains in the Rayleigh region, representing the low-magnetic-field operational domain where permeability is relatively constant.
    \item The secondary current is zero.
\end{itemize}

\subsection{Maxwell's Equations in 2D Plane}
We establish a local Cartesian coordinate system $(x, y, z)$ within the cross-section of the toroid. The origin $(x=0, y=0)$ is the center of the square cross-section. The $z$-axis of this local system runs tangentially along the mean circumference of the toroid.

Due to the problem's symmetry and the primary winding configuration, the magnetic field is purely axial (in the local $z$-direction), allowing $\mathbf{H} = H_z(x,y) \hat{\mathbf{z}}$. The induced electric field is purely in the $x-y$ plane. Assuming time-harmonic fields ($e^{j\omega t}$), Maxwell's equations inside the conductive core become:
\begin{align}
    \nabla \times \mathbf{H} &= \sigma \mathbf{E} \label{eq:ampere} \\
    \nabla \times \mathbf{E} &= -j\omega \mathbf{B} \\
    \mathbf{B} &= \mu_c \mathbf{H} \label{eq:faraday}
\end{align}

Taking the curl of Ampère's law and substituting Faraday's law yields the vector Helmholtz equation:
\begin{align}
    \nabla \times (\nabla \times \mathbf{H}) &= \sigma (\nabla \times \mathbf{E}) \\
    \nabla \times (\nabla \times \mathbf{H}) &= -j\omega \mu_c \sigma \mathbf{H}
\end{align}
Using the vector identity $\nabla \times (\nabla \times \mathbf{H}) = \nabla(\nabla \cdot \mathbf{H}) - \nabla^2 \mathbf{H}$ and knowing $\nabla \cdot \mathbf{H} = 0$, we obtain the 2D diffusion equation:
\begin{equation}
    \nabla^2_{xy} H_z - j\omega \mu_c \sigma H_z = 0
\end{equation}

By defining the complex propagation constant squared as $\gamma^2 = j\omega \mu_c \sigma$, the differential equation becomes:
\begin{equation}
    \frac{\partial^2 H_z}{\partial x^2} + \frac{\partial^2 H_z}{\partial y^2} = \gamma^2 H_z
\end{equation}

\subsection{Expressions for B and H}

To solve the partial differential equation, we use the separation of variables ansatz:
\begin{equation}
    H_z(x,y) = X(x)Y(y)
\end{equation}
Substituting this into the diffusion equation and dividing by $X(x)Y(y)$ yields:
\begin{equation}
    \frac{X''(x)}{X(x)} + \frac{Y''(y)}{Y(y)} = \gamma^2
\end{equation}
Since the square symmetry of the cross-section forces identical behavior along both the $x$ and $y$ axes, each term must equal a constant $k^2$ such that:
\begin{equation}
    2k^2 = \gamma^2
\end{equation}
\begin{align}
    k &= \sqrt{\frac{\gamma^2}{2}} \\
    k &= \sqrt{\frac{j\omega \mu_c \sigma}{2}}
\end{align}

The ordinary differential equations have general solutions built from hyperbolic functions. Constrained by the even symmetry of the core (the field must be symmetric about the $x$ and $y$ axes), the sinusoidal terms vanish, resulting in:
\begin{equation}
    H_z(x,y) = H_0 \cosh(kx)\cosh(ky)
\end{equation}

Applying Ampère's circuital law along the mean magnetic path $l_m = \pi D_m$, the field at the surface boundaries ($x = \pm a/2$ or $y = \pm a/2$) is established by the primary current:
\begin{equation}
    H_{surf} = \frac{N_1 I}{l_m}
\end{equation}
By evaluating the field at the center of a face, for instance at $(a/2, 0)$, we find the amplitude constant $H_0$:
\begin{align}
    H_0 \cosh(k a/2) \cosh(0) &= H_{surf} \\
    H_0 &= \frac{H_{surf}}{\cosh(k a/2)}
\end{align}

This gives the exact analytical expressions for $\mathbf{H}$ and $\mathbf{B}$:
\begin{equation}
    \boxed{H_z(x,y) = \frac{N_1 I}{l_m} \frac{\cosh(kx)\cosh(ky)}{\cosh(k a/2)}}
\end{equation}
\begin{align}
    B_z(x,y) &= \mu_c H_z(x,y) \\
    B_z(x,y) &= \mu_c \frac{N_1 I}{l_m} \frac{\cosh(kx)\cosh(ky)}{\cosh(k a/2)}
\end{align}
\begin{equation}
    \boxed{B_z(x,y) = \mu_c \frac{N_1 I}{l_m} \frac{\cosh(kx)\cosh(ky)}{\cosh(k a/2)}}
\end{equation}

\subsection{Flux Expression}
The total magnetic flux $\Phi$ passing through the square cross-section is the surface integral of $B_z(x,y)$:
\begin{align}
    \Phi &= \int_{-a/2}^{a/2} \int_{-a/2}^{a/2} B_z(x,y) \,dx \,dy \\
    \Phi &= \mu_c \frac{H_{surf}}{\cosh(k a/2)} \left( \int_{-a/2}^{a/2} \cosh(kx) \,dx \right) \left( \int_{-a/2}^{a/2} \cosh(ky) \,dy \right)
\end{align}
Evaluating the integrals yields:
\begin{align}
    \Phi &= \mu_c \frac{H_{surf}}{\cosh(k a/2)} \left( \frac{2}{k} \sinh(k a/2) \right)^2 \\
    \Phi &= \mu_c H_{surf} \frac{4}{k^2} \frac{\sinh^2(k a/2)}{\cosh(k a/2)}
\end{align}
By expanding $H_{surf}$, the final flux expression is framed as:
\begin{equation}
    \boxed{\Phi = \mu_c \frac{N_1 I}{l_m} \frac{4}{k^2} \frac{\sinh^2(k a/2)}{\cosh(k a/2)}}
\end{equation}

\subsection{Apparent Permeability}
If the core were a linear, non-conductive material without skin effect, the flux would be defined as $\Phi_{ideal} = \mu_{app} \frac{N_1 I A}{l_m} = \mu_{app} .H_{surf}. A$. Comparing this to the actual flux derived above, we define the complex apparent permeability under eddy current conditions:
\begin{align}
    \mu_{app} &= \frac{\Phi / a^2}{H_{surf}} \\
    \mu_{app} &= \mu_c \frac{4}{(ka)^2} \frac{\sinh^2(k a/2)}{\cosh(k a/2)}
\end{align}
\begin{equation}
    \boxed{\mu_{app} = \mu_c \frac{4}{(ka)^2} \frac{\sinh^2(k a/2)}{\cosh(k a/2)}}
\end{equation}

\subsection{Primary Complex Impedance Z}
By Faraday's law, the induced electromotive force in the primary is defined as $E_1 = j\omega N_1 \Phi$. The impedance of the core $Z_{core} = \frac{E_1}{I}$:
\begin{align}
    Z_{core} &= j\omega N_1 \frac{\Phi}{I} \\
    Z_{core} &= j\omega \mu_c \frac{N_1^2}{l_m} \frac{4}{k^2} \frac{\sinh^2(k a/2)}{\cosh(k a/2)}
\end{align}
By recalling the definition of the separation constant $k^2 = \frac{j\omega \mu_c \sigma}{2}$, we can substitute $\frac{j\omega \mu_c}{k^2} = \frac{2}{\sigma}$. This simplifies the core impedance to:
\begin{equation}
    \boxed{Z_{core} = \frac{8 N_1^2}{\sigma l_m} \frac{\sinh^2(k a/2)}{\cosh(k a/2)}}
\end{equation}

The primary winding wire also suffers from the skin effect. Let the wire have radius $r_w$, conductivity $\sigma_w$, and permeability $\mu_w$. The AC resistance $R_{winding}$ for circular wires is defined using Kelvin functions as described in \cite{tahaACloss}:
\begin{equation}
    R_{winding} = R_{DC} \frac{\xi}{2} \left[ \frac{\text{ber}(\xi) \text{bei}'(\xi) - \text{bei}(\xi) \text{ber}'(\xi)}{\text{ber}'^2(\xi) + \text{bei}'^2(\xi)} \right]
\end{equation}
where $\xi = \sqrt{2} r_w / \delta_w$. Including the leakage inductance $L_{leak}$, the total primary complex impedance is:
\begin{equation}
    \boxed{Z = R_{winding} + j\omega L_{leak} + \frac{8 N_1^2}{\sigma l_m} \frac{\sinh^2(k a/2)}{\cosh(k a/2)}}
\end{equation}

\subsection{Losses Expression}
The losses can be explicitly separated into Winding, Eddy Current, and Hysteresis losses using an admittance-based approach.

\textbf{1. Winding Loss:}
\begin{equation}
    \boxed{P_{w} = I_{rms}^2 R_{winding}}
\end{equation}

\textbf{2. Core Losses Separation:}
The core's admittance is computed as $Y_{core} = \frac{1}{Z_{core}} = G_{core} - jB_{core}$. The real part $G_{core}$ dictates the total active power lost in the core. To isolate the hysteresis loss, we analyze the hypothetical condition where the core's conductivity approaches zero ($\sigma \to 0$):
\begin{align}
    Z_h &= \lim_{\sigma \to 0} Z_{core} \\
    Z_h &= j\omega \frac{N_1^2 A}{l_m} \mu_c
\end{align}
The equivalent conductance responsible solely for intrinsic hysteresis is:
\begin{equation}
    G_h = \text{Re}\left( \frac{1}{Z_h} \right)
\end{equation}
The additional loss imposed by macroscopic induced currents characterizes the eddy current conductance:
\begin{equation}
    G_e = G_{core} - G_h
\end{equation}

Therefore, the explicitly separated core power losses become:

\textbf{Hysteresis Loss ($P_h$):}
\begin{equation}
    \boxed{P_h = |E_{1,rms}|^2 G_h}
\end{equation}

\textbf{Eddy Current Loss ($P_e$):}
\begin{equation}
    \boxed{P_e = |E_{1,rms}|^2 G_e}
\end{equation}
where $E_{1,rms} = |Z_{core}| I_{rms}$.

\subsection{Apparent Permeability}
To isolate the influence of eddy currents on the energy-storage capability of the core, the apparent permeability $\mu_{app}$ can be derived directly from the reactive component of the circuit. 

If we consider an equivalent idealized core where the macroscopic eddy currents are neglected and the inductive behavior is defined strictly by a real apparent permeability $\mu_{app}$, the purely inductive impedance is $Z_{ideal} = j\omega L_{app}$. Using the cross-sectional area $A = a^2$, the imaginary part of this idealized impedance is expressed as:
\begin{equation}
    \text{Im}(Z_{ideal}) = \omega \frac{N_1^2 a^2}{l_m} \mu_{app}
\end{equation}

Conversely, the exact analytical complex impedance of the core, which simultaneously accounts for the skin effect, eddy currents, and complex permeability, was derived as:
\begin{equation}
    Z_{core} = \frac{8 N_1^2}{\sigma l_m} \frac{\sinh^2(k a/2)}{\cosh(k a/2)}
\end{equation}
The imaginary part of this full complex impedance represents the true inductive reactance of the physical core:
\begin{equation}
    \text{Im}(Z_{core}) = \text{Im} \left( \frac{8 N_1^2}{\sigma l_m} \frac{\sinh^2(k a/2)}{\cosh(k a/2)} \right)
\end{equation}

By forcing the equivalent inductive model to match the true inductive behavior of the core, we equate the two imaginary components:
\begin{equation}
    \omega \frac{N_1^2 a^2}{l_m} \mu_{app} = \text{Im} \left( \frac{8 N_1^2}{\sigma l_m} \frac{\sinh^2(k a/2)}{\cosh(k a/2)} \right)
\end{equation}

Solving for $\mu_{app}$, the geometric and winding parameters $l_m$ and $N_1^2$ elegantly cancel out:
\begin{align}
    \mu_{app} &= \frac{l_m}{\omega N_1^2 a^2} \cdot \text{Im} \left( \frac{8 N_1^2}{\sigma l_m} \frac{\sinh^2(k a/2)}{\cosh(k a/2)} \right) \\
    \mu_{app} &= \frac{1}{\omega a^2} \text{Im} \left( \frac{8}{\sigma} \frac{\sinh^2(k a/2)}{\cosh(k a/2)} \right)
\end{align}

Since the angular frequency $\omega$, the dimension $a$, and the conductivity $\sigma$ are purely real and positive scalars, they can be consolidated inside the imaginary operator. This yields the final expression for the apparent permeability:
\begin{equation}
    \boxed{\mu_{app} = \text{Im} \left( \frac{8}{\omega \sigma a^2} \frac{\sinh^2(k a/2)}{\cosh(k a/2)} \right)}
\end{equation}

\section{Summary of Main Formulas}
For ease of reference, the fundamental analytical expressions determining the core characteristics derived in this paper are summarized in Table~\ref{tab:summary}.

\renewcommand{\arraystretch}{1.7}
\begin{table}[h]
\centering
\caption{Summary of Key Analytical Expressions}
\label{tab:summary}
\begin{tabular}{p{5.5cm} p{10cm}}
\toprule
\textbf{Quantity} & \textbf{Expression} \\
\midrule
Magnetic field $H_z(x,y)$ & $\displaystyle H_z(x,y) = \frac{N_1 I}{l_m} \frac{\cosh(kx)\cosh(ky)}{\cosh(k a/2)}$ \\[1em]
Flux density $B_z(x,y)$ & $\displaystyle B_z(x,y) = \mu_c \frac{N_1 I}{l_m} \frac{\cosh(kx)\cosh(ky)}{\cosh(k a/2)}$ \\[1em]
Magnetic flux $\Phi$ & $\displaystyle \Phi = \mu_c \frac{N_1 I}{l_m} \frac{4}{k^2} \frac{\sinh^2(k a/2)}{\cosh(k a/2)}$ \\[1em]
Apparent permeability $\mu_{app}$ & $\displaystyle \mu_{app} = \text{Im} \left( \frac{8}{\omega \sigma a^2} \frac{\sinh^2(k a/2)}{\cosh(k a/2)} \right)$ \\[1em]
Core impedance $Z_{core}$ & $\displaystyle Z_{core} = \frac{8 N_1^2}{\sigma l_m} \frac{\sinh^2(k a/2)}{\cosh(k a/2)}$ \\[1em]
Eddy current loss $P_e$ & $\displaystyle P_e = |E_{1,rms}|^2 (G_{core} - G_h)$ \\[1em]
Hysteresis loss $P_h$ & $\displaystyle P_h = |E_{1,rms}|^2 G_h$ \\[1em]
Winding AC resistance $R_{winding}$ & $\displaystyle R_{winding} = R_{DC} \frac{\xi}{2} \left[ \frac{\text{ber}(\xi) \text{bei}'(\xi) - \text{bei}(\xi) \text{ber}'(\xi)}{\text{ber}'^2(\xi) + \text{bei}'^2(\xi)} \right]$ \\[1em]
\bottomrule
\end{tabular}
\end{table}

\section{Results and Analysis}

The analytical models derived in the preceding sections are used to evaluate the electromagnetic characteristics of the square cross-section toroidal ring. The geometric, material, and excitation parameters used for the evaluation are summarized in Table~\ref{tab:sim_params}. The frequency response was evaluated across a logarithmic sweep from 10~Hz to 1~MHz using 500 sampling points.

\renewcommand{\arraystretch}{1}
\begin{table}[htbp]
    \centering
    \caption{Simulation and Material Parameters for Model Evaluation}
    \label{tab:sim_params}
    \begin{tabular}{lcc}
        \toprule
        \textbf{Parameter} & \textbf{Symbol} & \textbf{Value} \\
        \midrule
        \textbf{Geometry} & & \\
        Side length of square cross-section & $a$ & 5 mm \\
        Mean diameter of the ring & $D_m$ & 50 mm \\
        Number of primary turns & $N_1$ & 60 \\
        \midrule
        \textbf{Material Properties} & & \\
        Real relative permeability & $\mu_r'$ & 500 \\
        Hysteresis loss angle & $\theta$ & 0.1 rad \\
        Electrical conductivity & $\sigma$ & 900 S/m \\
        \midrule
        \textbf{Excitation} & & \\
        RMS primary current & $I_{rms}$ & 1.0 A \\
        Frequency range & $f$ & $10^1$ -- $10^6$ Hz \\
        \bottomrule
    \end{tabular}
\end{table}

\subsection{Spatial Distribution of Magnetic Flux Density}

\begin{figure}[htbp]
    \centering
    \begin{subfigure}[b]{0.45\textwidth}
        \centering
        \includegraphics[width=\textwidth]{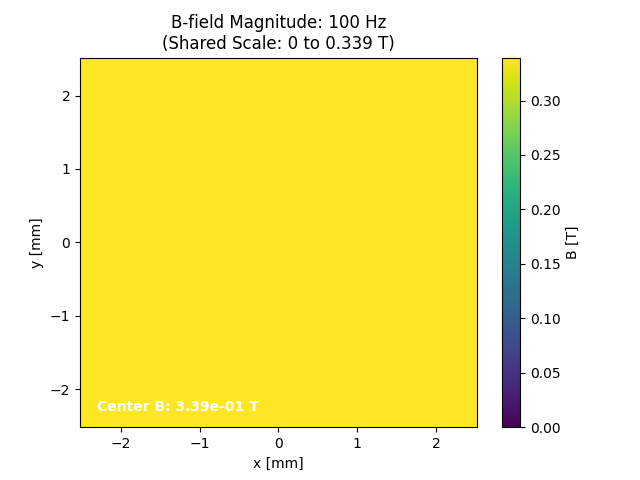}
        \caption{100 Hz: Quasi-static uniform distribution.}
        \label{fig:B100}
    \end{subfigure}
    \hfill
    \begin{subfigure}[b]{0.45\textwidth}
        \centering
        \includegraphics[width=\textwidth]{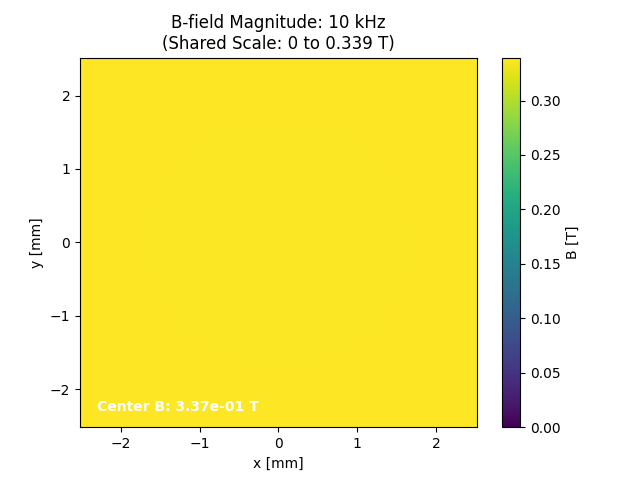}
        \caption{10 kHz: Onset of internal shielding.}
        \label{fig:B10k}
    \end{subfigure}
    
    \vspace{0.5cm}
    
    \begin{subfigure}[b]{0.45\textwidth}
        \centering
        \includegraphics[width=\textwidth]{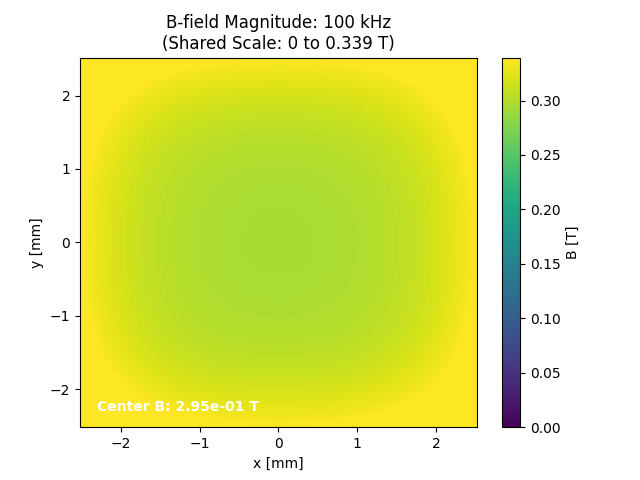}
        \caption{100 kHz: Pronounced flux crowding.}
        \label{fig:B100k}
    \end{subfigure}
    \hfill
    \begin{subfigure}[b]{0.45\textwidth}
        \centering
        \includegraphics[width=\textwidth]{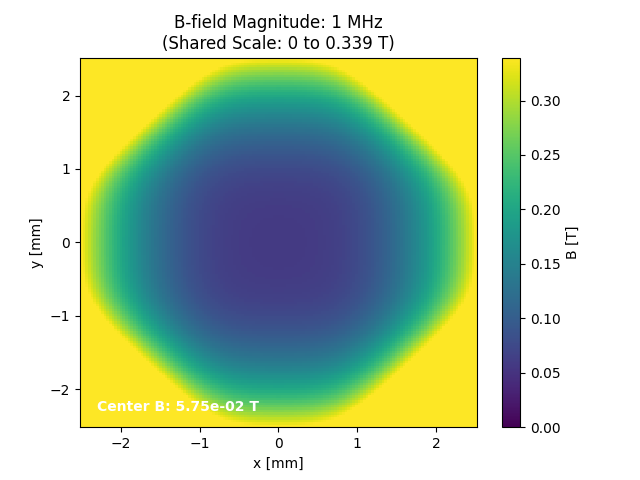}
        \caption{1 MHz: Extreme skin effect/expulsion.}
        \label{fig:B1M}
    \end{subfigure}
    \caption{Evolution of the Magnetic Flux Density magnitude across the square cross-section. The color scale is synchronized to highlight the collapse of internal flux as frequency increases.}
    \label{fig:B_profiles}
\end{figure}

Figure \ref{fig:B_profiles} illustrates the 2D spatial distribution of the magnetic flux density magnitude $|B_z(x,y)|$ across the square cross-section at four distinct frequencies.

At low frequency (100 Hz), the electromagnetic skin depth is significantly larger than the dimension $a$, resulting in a nearly uniform flux density of 0.339~T across the entire domain. As frequency increases to 100 kHz, the counter-magnetomotive force from induced eddy currents begins to repel the flux from the volumetric center, where $B$ drops to 0.295~T. At the 1 MHz limit, the skin effect is extreme; the magnetic field is virtually expelled from the core's interior, leaving a "dead zone" at the center ($B = 0.057$~T) and confining the flux to a thin peripheral layer.

\subsection{Apparent Permeability and Impedance Characteristics}

\begin{figure}[ht]
    \centering
    \includegraphics[width=0.65\textwidth]{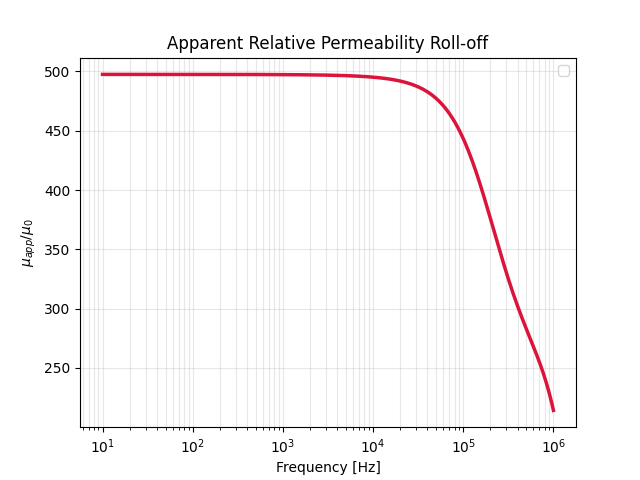}
    \caption{Apparent Relative Permeability roll-off showing the loss of inductive effectiveness at high frequencies.}
    \label{fig:mu_app}
\end{figure}

The macroscopic impact of the skin effect is captured by the apparent relative permeability $\mu_{app}/\mu_0$ (Fig. \ref{fig:mu_app}). The value remains constant at the intrinsic material permeability (500) until approximately 20 kHz. Beyond this threshold, the shielding effect causes a systematic roll-off, as the core can no longer effectively store magnetic energy within its volume.

\begin{figure}[ht]
    \centering
    \includegraphics[width=0.65\textwidth]{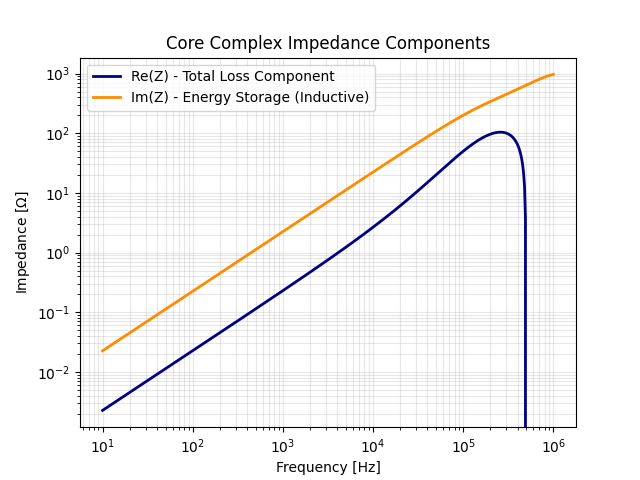}
    \caption{Primary complex impedance components showing the roll-off of both storage and loss mechanisms.}
    \label{fig:impedance}
\end{figure}

This roll-off is mirrored in the complex impedance components (Fig. \ref{fig:impedance}). While the inductive reactance $\text{Im}(Z)$ initially increases with frequency, it begins to deviate from linearity and eventually flattens. Simultaneously, the resistive component $\text{Re}(Z)$, representing total core losses, rises sharply before exhibiting a sharp attenuation at high frequencies.

\subsection{Loss Separation and Eddy Current Roll-off}

\begin{figure}[ht]
    \centering
    \includegraphics[width=0.65\textwidth]{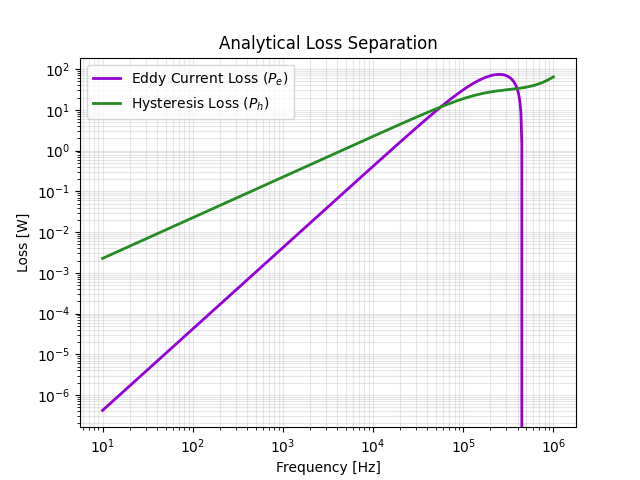}
    \caption{Analytical separation of losses. Note the characteristic "crash" in eddy current power dissipation at high frequencies.}
    \label{fig:losses}
\end{figure}

Figure \ref{fig:losses} provides the analytical separation of losses. Hysteresis loss ($P_h$) scales linearly with frequency at low excitation levels but experiences attenuation as the active magnetic volume shrinks. 

The most notable observation is the behavior of the eddy current loss ($P_e$). Initially, $P_e$ grows with an $\omega^2$ proportionality. However, after approximately 300 kHz, the eddy current loss reaches a peak and then decreases sharply. 

This roll-off is a result of inductive shielding. The power loss is proportional to the square of the induced EMF ($P_e \propto E^2$), and $E$ is proportional to the product of angular frequency and total magnetic flux ($\omega \Phi$). At low frequencies, $\Phi$ is constant, so $P_e$ rises with $\omega^2$. However, at very high frequencies, the skin effect becomes so severe that the total flux $\Phi$ inside the conductive core collapses faster than the frequency $\omega$ increases. Effectively, the core behaves as a magnetic shield; the field is reflected at the surface rather than penetrating the material. As the flux $\Phi$ approaches zero, the mechanisms for internal dissipation vanish, leading to the observed decrease in $P_e$. It is important to emphasize that the results presented in this study assume a constant-current excitation across the entire frequency range. This differs from the standardized measurement systems, such as those by Brockhaus or Iwatsu, which typically prioritize maintaining a constant peak magnetic flux density. In this case, the primary current is increased at higher frequencies to compensate for the flux reduction induced by the skin effect and internal magnetic shielding.

\clearpage
\bibliographystyle{ieeetr}
\bibliography{references}

\end{document}